\documentclass{mem}
\usepackage{natbib}\usepackage{txfonts}\usepackage{balance}
\usepackage{graphicx}
\usepackage[a4paper,breaklinks,dvipdfm]{hyperref}
\idline{75}{282}
\begin{document}

\title{
The SONYC survey: Towards a complete census of brown dwarfs in star forming regions
}

\author{
K. \,Mu\v{z}i\'{c}\inst{1}, 
A. \,Scholz\inst{2},
V. \,C. Geers\inst{3},
R. \,Jayawardhana\inst{4},
M. \,Tamura\inst{5},
P. \,Dawson\inst{3},
\and T. P. \,Ray\inst{3}
}

\offprints{K. Muzic}

\institute{
European Southern Observatory  --
Alonso de C\'ordova 3107, Casilla 19, 
Santiago, Chile
\and
School of Physics \& Astronomy, St. Andrews University --
North Haugh, St Andrews KY16 9SS, United Kingdom
\and
School of Cosmic Physics, Dublin Institute for Advanced Studies -- 
31 Fitzwilliam Place, Dublin 2, Ireland
\and
Department of Astronomy \& Astrophysics, University of Toronto -- 
50 St. George Street, 
Toronto, ON M5S 3H4, Canada
\and
National Astronomical Observatory --
Osawa 2-21-2, Mitaka, Tokyo 181, Japan\\
\email{kmuzic@eso.org}
}

\authorrunning{Muzic}
 
\titlerunning{SONYC}

\abstract{
SONYC, short for "Substellar Objects in Nearby Young Clusters", is a survey program to 
provide a census of the substellar population in nearby star forming regions. 
We have conducted deep 
optical and near-infrared photometry in five young regions (NGC1333, $\rho$ Ophiuchi, Chamaeleon-I, Upper Sco, and
Lupus-3), combined with proper motions, 
and followed by extensive spectroscopic campaigns with Subaru and VLT, in which 
we have obtained more than 700 spectra of candidate low-mass objects.
We have identified
and characterized more than 60 new substellar objects, among them a handful
of objects with masses close to, or below the Deuterium burning limit.
Through SONYC and surveys by other groups, the substellar IMF is now well characterized 
down to $\sim5 - 10 $M$_{Jup}$, and we find that the ratio of the number of stars
with respect to brown dwarfs lies between 2 and 6. 
A comprehensive survey of NGC~1333 reveals that, down to $\sim5$M$_{Jup}$, free-floating objects with planetary masses
are 20-50 times less numerous than stars, i.e. their total contribution to the mass budget
of the clusters can be neglected. 
}
\maketitle{}

\section{Introduction}

The low-mass end of the Initial Mass Function (IMF) has been the subject of intensive investigation
over almost two decades, but the phyical processes that set an object's mass in this regime
are still matters of debate (e.g. \citealt{bonnell07, bastian10, jeffries12}). 
Brown dwarfs (BDs) could in theory be produced by various processes, which include
dynamical ejections from multiple systems or
disks, gravitational fragmentation,
and photoerosion of cores in the vicinity of OB stars \citep{whitworth07}, but 
their relative importance is not yet clear. Also,
deep surveys in star forming regions have revealed a population of objects
the masses of which are estimated to fall below the Deuterium-burning limit ($\sim$12 M$_{Jup}$), i.e.
there seems to be an overlap in the mass space between free-floating objects in clusters, and giant
planets. Detailed studies of the substellar mass regime at young ages 
are therefore crucial to understand the mass dependence in the formation and early evolution of stars and planets. 

SONYC - short for {\it Substellar Objects in Nearby Young Clusters} - is a comprehensive project aiming
to provide a complete census of substellar population down to a few Jupiter masses
in young star forming regions. By identifying large, unbiased, and well-characterized samples of brown dwarfs in various regions,
we can provide the best possible candidates for detailed follow-up studies (e.g. disks, accretion,
multiplicity), and
try to investigate possible environmental effects. In this contribution we give an overview of all SONYC efforts to date, 
and summarize the most important results.

\section{The SONYC survey}

\begin{figure}[t!]
\resizebox{\hsize}{!}{\includegraphics[clip=true]{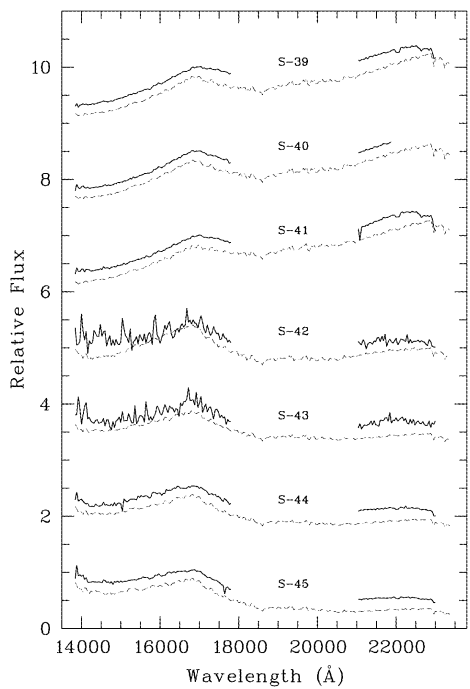}}
\caption{\footnotesize
Near-infrared spectra of the 7 identified very low
mass members of NGC1333 (solid lines). The dashed lines show
the best-fitting, reddened model spectra. 
The spectra are offset on the y-axis by constants for clarity. Figure 
from \citet{scholz12b}.
}
\label{spec}
\end{figure}

SONYC is designed to reach mass limits well below 0.01$\,$M$_{\odot}$, with the main 
candidate selection method based on the optical photometry. By detecting the photosphere 
we ensure we obtain a realistic picture of the substellar population in each of the studied clusters, 
avoiding the biases introduced by the mid-infrared selection 
(only objects with disks), or methane-imaging (only T-dwarfs). 
For the detection of low-mass, pre-main sequence sources, we use
our own extremely deep broad-band optical- and near-infrared imaging, in combination with the public photometry catalogues, such as 
the Two Micron All Sky Survey (2MASS) and the United Kingdom Infrared Digital Sky Survey (UKIDSS). 
Broad-band selection in regions with significant extinction results
in candidate lists with plenty of contamination by embedded stellar members, 
reddened background M-type stars, and (less likely) late M- and early L-type objects in the foreground. Follow-up spectroscopy is therefore
mandatory to confirm the photometric candidates as young very low-mass (VLM) objects. Since in some cases our
candidate lists may contain hundreds of objects, the only way to obtain spectra for a large fraction of a sample is by means
of multi-objects spectrographs. For this purpose we successfully used MOIRCS \citep{ichikawa06} and FMOS \citep{kimura10} at the Subaru Telescope, and VIMOS \citep{lefevre03} at the ESO's VLT.
Other facilities used for the follow-up include SINFONI \citep{eisenhauer03} at the VLT, and SpeX at the IRTF \citep{rayner03}.
Proper motion analysis can greatly facilitate the candidate selection, and was applied to two of the regions studied so far, Upper Sco and Lupus~3. 
In Fig.~\ref{spec} we show MOIRCS spectra of 7 confirmed VLMOs from the latest observing campaign in NGC~1333. Dashed lines show AMES-DUSTY model
spectra \citep{allard01} which are used to estimate effective temperatures. As evidence for membership in the near-infrared we use the 
triangular shape of the water absorption feature in the H-band, typical for young objects later than M3 \citep{cushing05}. For the spectra taken at optical
wavelengths, we identified M-type objects based on the characteristic molecular bands (mainly TiO and VO; \citealt{kirkpatrick91}), and use
several gravity-dependent spectral features, such as Na~I absorption at $\sim 8200$\AA. 

In the framework of SONYC we have carried out imaging and spectroscopy in five regions (NGC~1333, $\rho$-Ophiuchus, Chamaeleon-I, Lupus~3, and Upper Scorpius).
Our results in Cha~I indicate that the current census from \citet{luhman07} is
mostly complete down to $\sim$8 M$_{Jup}$ and A$_V \leq\,$5 mag \citep{muzic11}. 
In $\rho$-Oph we find 10 new VLM sources, and 
estimate that the number of missing substellar objects in our survey area
centered on the main core L1688 is $\sim 15$, down to $0.003 - 0.03 M_{\odot}$  and for $A_V = 0 - 15$
\citep{geers11,muzic12}.  

\begin{figure}[t!]
\resizebox{\hsize}{!}{\includegraphics[clip=true]{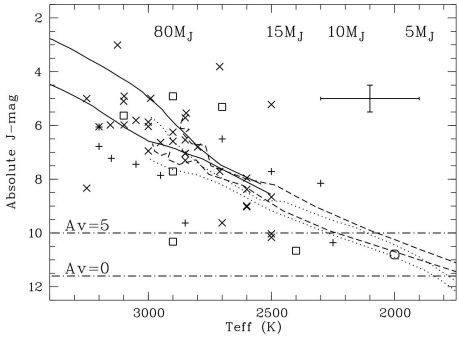}}
\caption{\footnotesize
HR diagram for the very low mass population in NGC1333
(crosses -- objects known before in the census of \citet{scholz12a},
pluses -- new objects found in the latter study, squares -- new objects
identified in \citet{scholz12b}). The circle shows the improved parameters
for SONYC-NGC1333-36 after fitting the photometric SED. 
Model isochrones (solid -- BCAH, dotted -- DUSTY,
dashed-- COND) are shown for 1 (upper) and 5 (lower) Myr. The
dash-dotted lines show the limits of our spectroscopic survey for
A$_V$ = 0 and 5mag.
}
\label{HRD}
\end{figure}

In NGC~1333, a compact cluster in Perseus, we obtained spectra of $\sim 80 \%$ of all the photometric candidates, making this cluster
one of the best studied examples in this scientific field \citep{scholz09, scholz12a, scholz12b}. Among the population of 30-40 spectroscopically
confirmed BDs, we find three objects with spectral type $\geq\,$M9, i.e. with masses probably lying in the planetary domain. It is in this
cluster that we have identified the coolest SONYC brown dwarf. With a spectral type of $\sim$L3, its estimated mass is about $\sim 0.006\,$M$_{\odot}$, according 
to the COND \citep{baraffe03} and DUSTY \citep{chabrier00} models. 
From the HR-diagram for the very low mass population in NGC~1333 shown in Figure ~\ref{HRD}, it 
is evident that there is a drop in frequency of observed objects at effective temperatures below 2500 K. As
discussed in detail in \citet{scholz12b}, possible biases of our photometric selection cannot have a major influence
on this result, i.e. the drop in the number of the objects in the lowest mass bin appears to be real. 
This indicates that, down to $\sim5$M$_{Jup}$, the free-floating objects with planetary masses in NGC~1333
are 20-50 times less numerous than stars. 

In a project complementary to SONYC, we have made use of data from the Galactic Cluster Survey, a part of
UKIDSS, to search for brown dwarfs in the Upper Scorpius OB association. From the list of high-confidence 
candidates selected from photometry and proper motions, we obtained 30 spectra and confirmed 24 young very-low mass members of UpSco 
(\citealt{dawson11}, Dawson et al., in prep). Of these 24, 15-20 are likely to be BDs with masses between 0.01 to 0.08 M$_{\odot}$

In Table~\ref{table} we give an overview of the surveyed regions, along with their approximate age and distance, and the area covered by each photometric survey. 
We also give the completeness levels translated to object mass, by comparison with COND \citep{baraffe03}, DUSTY \citep{chabrier00} and BT-Settl \citep{allard11} models.
We list the total number of photometric candidates, the number of spectra obtained in the follow-up, and the approximate number of confirmed substellar objects. 
The main candidate selection is based on optical and near-infrared photometry, and the number of candidates from this selection is given for all the regions. In two cases we also give
the number of candidates selected from {\it Spitzer} mid-infrared data (second number following the plus sign).

\begin{table*}
\caption{Summary of the SONYC survey}
\label{table}
\begin{center}
\begin{tabular}{lccccc}
\hline
\\
 & Cha~I & $\rho$~Oph & NGC~1333 & Lupus~3 & Up~Sco\\
\hline
\\
age [Myr]	& 2    & 1    & 1   & 1   & 5-10 \\
d [pc]		& 160  & 125  & 300 & 200 & 145  \\
area [deg$^2$]	& 0.25 & 0.33 & 0.25 & 1.4 & 57   \\
\hline
completeness [M$_{\sun}$] & 0.008 & 0.003 -- 0.03 & 0.004 -- 0.008 & 0.009 -- 0.02 & 0.02 \\
at A$_V$	& $\leq5$ & 0--15 & 0 -- 5 & 0 -- 5 & $<$5 \\
\hline
\# of candidates & 142 & 309 + 83 & 196  + 10 & 409 & 96 \\
\# of spectra	 & 60 & 160 & 160 & 124 & 30\\
\# of BDs	 & $\sim$9 & $\sim$15 & 30-40 & $\sim$4 & 15-20
\\
\hline
\end{tabular}
\end{center}
\end{table*}


\section{Brown dwarf frequencies}

The ratio of the number of stars to that of BDs has
been often used to compare the abundance of substellar objects in different clusters and star forming regions.
Here, for the sake of completeness, the low-mass cutoff of the bin representing BDs is set to 0.03 M$_{\odot}$, while the stellar mass bin
is set to masses between 0.08 to 1.0 M$_{\odot}$. By compiling the available numbers from the
literature, in \citet{scholz12a} we have presented tentative evidence for regional differences
in the star/BD ratio. The published values range from 2 to 8. At the two extreme values of this range we find two
young clusters in the Perseus star forming region, NGC~1333 and IC~348. In NGC~1333, our survey led to a star/BD ratio of 2,
while \citet{luhman03} and \citet{andersen08} find a ratio of 8 for the slightly older IC~348 cluster. While at face value this 
might indicate variations in the BD abundance by a factor of 4, the uncertainties in these numbers have not been thoroughly investigated. 
In our most recent SONYC contribution (Scholz et al., 2013, submitted), we revisit the issue by deriving the star/BD ratio in a consistent manner for the two 
clusters, and investigate the effects that different assumptions imprint on the results. An important part
of this work is a comprehensive evaluation of uncertainties due to sample sizes, isochrones used to derive masses, and cluster parameters such as the assumed distance and age.
For NGC~1333 we a find star/BD ratio between 1.9 and 2.4, consistent with our previous work. For IC~348, the ratio is found to be between 2.9 and 4.0, suggesting
that the value has been overestimated by the previous studies. For these two samples, we find a typical error of $\pm$1 in star/BD ratios, which can
be lowered once we have more accurate age and distance estimates.
We find that the star forming process generates about 2.5 -- 5 substellar objects per 10 stars, which corresponds to a slope
of the power-law mass function of $\alpha$= 0.7 -- 1.0.

\section{Summary}

SONYC has been an ongoing survey program to provide a substellar population census in
nearby star forming regions. Here we summarize the main achievements of the survey:\\ 
\\
1. The discovery and characterization of more than 60 new substellar objects,
including a handful of objects with masses close to, or below the
Deuterium burning limit.\\
\\
2. Through SONYC and similar surveys by other groups, the substellar IMF
is now well characterized down to 5-10 M$_{Jup}$, with the number of BDs per
10 stars constrained to 2.5-5 in clusters. This is presently in
disagreement with the results from the field population, where fewer BDs
per star are found.\\
\\
3. Down to $\sim$5$\,$M$_{Jup}$, free-floating planetary mass objects are observed to
be rare, 20-50 times less numerous than stars. This implies that their
total contribution to the mass budget of the clusters can be neglected.
This is in clear disagreement with results of microlensing surveys \citep{sumi11}, which state that
free-floating planetary-mass objects are twice as common as stars (see also contribution by J. P. Beaulieu in the current proceedings), and with
the study in $\sigma$Ori which claims a turnover in the mass function around Deuterium-burning mass limit (contribution of M. R. Zapatero Osorio).\\
\\
4. One of the goals of SONYC is to provide samples of VLM objects that can be used 
by the community for different follow-up studies. All the published spectra and photometric catalogs
will be available online ({\it http://browndwarfs.org/sonyc}).
 
\begin{acknowledgements}
This work was co-funded under the Marie Curie Actions of the European
Commission (FP7KCOFUND). 
A.S., P.D., and TPR would like to acknowledge Science Foundation Ireland support through grant number 10/RFP/AST278.
\end{acknowledgements}

\bibliographystyle{aa}

\end{document}